\documentclass[aps,pra,superscriptaddress,showpacs]{revtex4}
\usepackage[intlimits]{amsmath}
\usepackage{amsfonts}
\usepackage{verbatim} 
\usepackage{color}
\usepackage{ifpdf}

\ifpdf
\usepackage[pdftex]{hyperref}
\else
\usepackage[hypertex,ps2pdf,dvips,colorlinks,linkcolor=red,urlcolor=blue,citecolor=blue]{hyperref}
\fi

\advance\voffset by -1cm
\advance\hoffset by -0.5cm
\newcommand{\be}[1]{\begin{equation}\label{#1}}
\newcommand\ee            {\end{equation}}
\newcommand\bes           {\begin{subequations}}
\newcommand\esu           {\end{subequations}}
\newcommand\erf[1]        {\eqref{#1}}

\newcommand{\ud}{\mathrm d}

\renewcommand\l             {\lambda}
\newcommand\mc            {\mathcal}

\newcommand\sg           {Sinh--Gordon }

\newcommand\p            {\partial}

\newcommand\psid         {\psi^{\dagger}}
\renewcommand\th         {\theta}

\renewcommand\vec[1]{{\vert{#1}\rangle}}
\newcommand\cev[1]{{\langle{#1}\vert}}
\newcommand\vac{{\vec 0}}
\newcommand\cav{{\cev 0}}

\newcommand\FF[1]{{\langle0\vert#1\vert\th_1,\dots,\th_n\rangle}}
\newcommand\no[1]{{\,:\!#1\!:\,}}
\def\ceev#1#2{{\langle{#1}\vert{#2}\rangle}}
\def\3pt#1#2#3{{\langle{#1}\vert{#2}\vert{#3}\rangle}}

\newcommand\arxiv[2]      {\href{http://arXiv.org/abs/#1}{#2}}
\newcommand\doi[2]        {\href{http://dx.doi.org/#1}{#2}}
\newcommand\httpurl[2]    {\href{#1}{#2}}

\begin{document}

\title{Bethe Ansatz Matrix Elements as Non-Relativistic Limits \\
of  Form Factors of Quantum Field Theory}

\author{M. Kormos}
\affiliation{SISSA and INFN, Sezione di Trieste, via Beirut 2/4, I-34151, 
Trieste, Italy}

\author{G. Mussardo}
\affiliation{SISSA and INFN, Sezione di Trieste, via Beirut 2/4, I-34151, 
Trieste, Italy}
\affiliation{International Centre for Theoretical Physics (ICTP), 
I-34151, Trieste, Italy}

\author{B. Pozsgay}
\affiliation{Institute for Theoretical Physics, Universiteit van Amsterdam,\\
Valckenierstraat 65, 1018 XE Amsterdam, The Netherlands}

\begin{abstract}
\noindent
We show that the matrix elements of integrable models computed by the
Algebraic Bethe Ansatz can be put in direct correspondence with the
Form Factors of integrable relativistic field theories. This happens
when the $S$-matrix of a Bethe Ansatz model can be regarded as a
suitable non-relativistic limit of the $S$-matrix of a field theory,
and when there is a well-defined mapping between the Hilbert spaces
and operators of the two theories. This correspondence provides an
efficient method to compute matrix elements of Bethe Ansatz integrable
models, overpassing the technical difficulties of their direct
determination. We analyze this correspondence for the simplest example
in which it occurs, i.e. the Quantum Non-Linear Schr\"odinger and the
Sinh--Gordon models.
\end{abstract}


\maketitle

\section{Introduction}

The aim of this paper is to show that a direct correspondence exists
between the matrix elements \cite{footnote1} computed by the Algebraic
Bethe Ansatz in non-relativistic integrable models (in the following
simply referred to as Bethe Ansatz models) \cite{korbook} and the Form
Factors considered in relativistic integrable quantum field theories
\cite{smirnov}. As shown below, the relation between the two
quantities can be established along the lines of the recent studies on
non-relativistic limit of a quantum field theory \cite{KMT1,KMT2}.

The discovery of such a correspondence may greatly help deepen our
general knowledge of integrable models and, in particular, shed new
light on the calculation of their correlation functions. The reason is
the following: while the direct computation of Bethe Ansatz matrix
elements proves to be quite a difficult task (often carried out
successfully only for few operators), the computation of Form Factors
is instead a simpler type of problem. In the latter case, for instance, one
can take advantage of additional constraints coming from the
relativistic invariance of the field theory and as a result, explicit
expressions of Form Factors can be usually found not only for a few
operators but also for a larger number of them: actually, the
classification of all operators of a quantum field theory can be
obtained in terms of the different solutions of the Form Factor
equations \cite{cardymussardo,delfino}.

At the heart of this correspondence lies the $S$-matrix, both of the
relativistic field theory and the Bethe Ansatz model: if the
$S$-matrix of the latter is obtained by a suitable non-relativistic
limit of the $S$-matrix of the former, then the Form Factors of the
quantum field theory go smoothly to the Bethe Ansatz matrix
elements. Obviously one has to be sure that there is also a one-to-one
mapping between the Hilbert spaces and operators of the two
theories. But if one can prove that such a mapping exists, it is then
easy to understand why the field theory Form Factors reduce to the
Bethe Ansatz matrix elements: this happens because the analytic
properties of the Form Factors and the Bethe Ansatz matrix elements
are dictated by the $S$-matrices of the corresponding theories. In the
following we provide evidence for this correspondence by analyzing the
simplest models in which it occurs: the Quantum Non-Linear
Schr\"odinger (QNLS) model on one side, and the Sinh--Gordon (Sh-G)
model on the other. In particular, our approach provides a universal
method to compute matrix elements of any local operator in the QNLS
model.



\section{Matrix elements in the Algebraic Bethe Ansatz}

In this section we summarize without derivation some basic properties
and results of the Algebraic Bethe Ansatz solution of the QNLS system
associated to the Lieb--Liniger model \cite{LL}. The interested reader
is referred to the book \cite{korbook} and references therein.

The Hamiltonan of the model in volume $L$ with periodic boundary
conditions is given by
\be{HLL}
H_{\text{QNLS}}=\int_{0}^{L}\,\ud x\left(\p_x\psid\p_x\psi+ c\psid\psid\psi\psi\right)\;,
\ee
where $\psi(x,t)$ and $\psi^\dagger(x,t)$ are canonical Bose
fields
\be{Psi}
[\psi(x,t),\psi^\dagger(y,t)]=\delta(x-y)\;,
\ee
and  $c$ is the coupling constant. The Fock vacuum is defined as 
\be{fock}
\psi\vac=0,\qquad\cav\psid=0\;.
\ee
The QNLS model can be solved via the Algebraic Bethe Ansatz (BA).  
The monodromy matrix reads 
\be{mon}
T(\l)=\left(
\begin{array}{cc} A(\l)&B(\l)\\C(\l)&D(\l)
\end{array}\right)
\ee
and its entries act in a space consisting of states
\be{ABA} 
\vec{\l_1,\dots,\l_N}=\prod_{j=1}^N \mathbb{B}(\l_j)\vac\;,\qquad N=0,1,\dots\;.
\ee
where $\mathbb{B}(\l)=B(\l)\exp(-i\l L/2)$, $\{\l\}$ are arbitrary
complex parameters, and the pseudo-vacuum $\vac$ coincides with the
Fock-vacuum. Similarly, dual states can be constructed using the
operators $\mathbb{C}(\l)=C(\l)\exp(-i\l L/2)$.

The $R$-matrix describes the commutation relations of the monodromy
matrix entries and it satisfies the Yang--Baxter equations. For the
QNLS model it can be written in the form
\be{Rm}
R(\l,\mu)=\left(
\begin{array}{cccc}
f(\mu,\l) &0&0&0\\
0&g(\mu,\l)&1&0\\
0&1&g(\mu,\l)&0\\
0&0&0&f(\mu,\l)
\end{array}\right)
\ee
with
\be{fg} 
f(\mu,\l)=\frac{\mu-\l+ic}{\mu-\l}\;,\qquad g(\mu,\l)=\frac{ic}{\mu-\l}\;.
\ee

The transfer matrix $\tau(\l)=\mathop{\rm tr} T(\l) =A(\l)+D(\l)$
generates the complete set of the conservation laws of the model. The
eigenstates of the transfer matrix have the form \erf{ABA}, however
the parameters $\{\l\}$ are not arbitrary but they satisfy the system
of Bethe equations
\be{BY} 
e^{i\l_j L}\prod_{k=1\atop{k\ne j}}^N \tilde S_{\text{QNLS}}(\l_j,\l_k) = 1\;, 
\ee
where the two-particle $S$-matrix is given by
\be{SLL}
\tilde S_{\text{QNLS}}(\l_j,\l_k)\equiv\frac {f(\l_k,\l_j)}{f(\l_j,\l_k)}=
\frac{\l_j-\l_k-ic}{\l_j-\l_k+ic}\;.
\ee
The $S$-matrix gives the phase factor by which the state gets
multiplied when the particles $i$ and $j$ are interchanged. Hence the
Bethe equations say that the total phase-shift acquired when a
particle of momentum $\l_j$ is taken to a round trip comes from the
usual phase which is proportional to the momentum plus the scattering
phase shifts picked up when the particle scatters through all the
other particles. Taking the logarithm leads us to
\be{BYlog}
\tilde Q_j
= \l_jL+\sum_{k\ne j}^N\frac1i\log \tilde S_{\text{QNLS}}(\l_j,\l_k)=
2\pi I_j\;,\qquad I_j\in\mathbb{Z}\;.
\ee

Using the algebra satisfied by the monodromy matrix, the scalar
products of the BA states \erf{ABA} can be worked out explicitly, as
well as the action of the operator $\psi$ on these states. However,
the calculation of the scalar products proved to be a highly
non-trivial combinatorial problem (see \cite{korbook} and references
therein).  As a result, the norms of the states with parameters $\l$
that satisfy the Bethe equations \erf{BY} are
\be{norm}
\ceev{\l_1,\dots,\l_N}{\l_1,\dots,\l_N} = c^N
\left(\prod_{j,k=1\atop{j\ne k}}^Nf(\l_j,\l_k)\right)\tilde\rho_N\;,
\ee
where 
\be{rhoNR}
\tilde\rho_N=\det\left(\frac{\p \tilde Q_j}{\p\l_k}\right)
\ee
is the Gaudin determinant associated to the Bethe equations
  \erf{BY}.
Knowing the action of $\psi$ on the BA states and the scalar products,
its {\it unnormalized} matrix elements  
\be{FF}
\tilde F_N^{\psi}(\l'_1,\dots,\l'_{N-1}|\l_1,\dots,\l_N) = \3pt{\l'_1,\dots,\l'_{N-1}}{\psi(0,0)}{\l_1,\dots,\l_N}
\ee
can be given explicitly. These matrix elements are originally defined
for states which solve the Bethe equations. However, one can define a
function $F_N$ such that the actual matrix elements will be given by
the value of this function taken at the particular set of momenta
which satisfy the Bethe equations. Hence, the function $F_N$ itself
does not carry any information about the system size $L$: in fact,
this only enters the Bethe equations satisfied by the physical
momenta. Note that the only non-zero matrix elements of $\psi$ are for
states where the difference of the particle numbers is one and that
the functions $F_N$ are symmetric separately in the momenta $\l$ and
$\l'$.
We will give explicit examples for matrix elements in section
\ref{last}. However, it is important to note that they satisfy the recursion
relation \cite{izkorresh}
\begin{multline}
\label{recurs}
\tilde F_N^{\psi}(\l'_1,\dots,\l'_{N-1}|\l_1,\dots,\l_N)\xrightarrow[\l'_1\to\l_1]{} \\
\frac{ic}{\l_1-\l'_1}\left(\prod_{j=2}^{N-1}f(\l'_1,\l'_j)\prod_{j=2}^{N}f(\l_j,\l_1)
- \prod_{j=2}^{N-1}f(\l'_j,\l'_1)\prod_{j=2}^{N}f(\l_1,\l_j)\right)\;\times\\
\times\;\tilde F_{N-1}^{\psi}(\l'_1,\dots,\l'_{N-2}|\l_1,\dots,\l_{N-1}) + \dots\;,
\end{multline}
where the dots stand for non-singular parts. 

Similarly, matrix elements of the current operator
$j(x)=\psid(x)\psi(x)$ can also be determined. The current has
non-zero matrix elements only between states having the same number of
particles,
\be{FFj}
\tilde F_N^{j}(\l'_1,\dots,\l'_N|\l_1,\dots,\l_N) = \3pt{\l'_1,\dots,\l'_N}{\psid(0,0)\psi(0,0)}{\l_1,\dots,\l_N}\;,
\ee
and they also satisfy the recursion relations \erf{recurs} with the
obvious change in the number of particles of the dual vector $(N-1)\to
N$.

\section{Form Factors in the \sg model}

The Sh-G model is an integrable relativistic field
theory in $1+1$ dimensions defined by the Lagrangian density
\be{shgL}
\mc{L}= \frac12\left(\frac{\p\phi}{c_l\,\p
  t}\right)^2-\frac12\left(\frac{\p\phi}{\p x}\right)^2 -
\frac{m_0^2c_l^2}{g^2\hbar^2}\left(\cosh(g\phi)-1\right)\,, 
\ee
where $\phi=\phi(x,t)$ is a {\em real} scalar field, $m_0$ is a mass
scale and $c_l$ is the speed of light. The parameter $m_0$ is related to
the physical (renormalized) mass $m$ of the particle by \cite{babkar} 
\be{m0}
m_0^2\,=\,m^2\frac{\pi\alpha}{\sin(\pi\alpha)}\,.
\ee
The integrability of the Sh-G model implies the absence of particle
production processes and its $n$-particle scattering amplitudes are
purely elastic. Moreover, they factorize into $n(n-1)/2$ two-body
$S$-matrices which can be determined exactly via the $S$-matrix
bootstrap \cite{zamzam}. The energy $E(\th)$ and the momentum $P(\th)$
of a particle can be written as $E(\th)=M c_l^2 \cosh\th$, $P(\th)=M
c_l \sinh\th$, where $\th$ is the rapidity. In terms of the rapidities
the exact two-body $S$-matrix is given by \cite{ari}
\be{shgSmat}
S_{\text{Sh-G}}(\th_1,\th_2)\equiv S(\th_1,\th_2)=
\frac{\sinh\th_{12}-i\,\sin(\alpha\pi)}{\sinh\th_{12}+i\,\sin(\alpha\pi)}\;,
\ee
where $\th_{12}=\th_1-\th_2$ and $\alpha$ is the renormalized coupling constant 
\be{alpha}
\alpha\,=\,\frac{\hbar c_l\,g^2}{8\pi+\hbar c_l\,g^2}\;.
\ee

The key observation, made in papers \cite{KMT1,KMT2}, is that the
QNLS model can be regarded as a suitable non-relativistic limit of the
Sh-G model, under which the two-particle $S$-matrices, the Hamiltonian
and the Thermodynamic Bethe Ansatz equations of the Sh-G model go into
the corresponding quantities of the QNLS model. The connection between
the two theories is realized by taking a double limit, in which the
speed of light $c_l$ goes to infinity, the coupling $g$ goes to zero,
but with their product kept fixed and given by
\be{lim}
c_l\to\infty\,\,\, ,\,\,\, g\to0\;,\;\quad g\,c_l=\frac{4\sqrt{c}}\hbar\;,
\ee
where $c$ is the coupling constant of the QNLS model. Taking such a
double limit of the $S$-matrix of the Sh-G model one arrives at the
$S$-matrix \erf{SLL} of the QNLS model, once we set $m=1/2$ and
$\hbar=1$. Note that $m_0\to m$ in the limit.

Such a correspondence between $S$-matrices gives a hint that an exact
mapping exists between the two models. Given that in a relativistic
integrable model the two-particle $S$-matrix governs its entire
dynamics (its thermodynamic properties and the Form Factors of all its
operators), if this mapping exists then it has several interesting and
far-reaching consequences. For instance, in the past a mapping
  between two relativistic $S$-matrices was used to establish the
  correspondence between the form factors and, based on this, between
  the operators of the theories \cite{ahnmusdel}. As discussed below,
  in the present case the situation is more subtle, because the
  mapping operates between a {\em relativistic} theory and {\em a
    non-relativistic} model. The analysis in this case requires
  additional care about the structure of states and operators of the
  two Hilbert spaces.
 
Here we would like to shed light on the direct relation between the
Bethe Ansatz matrix elements of the QNLS and the Form Factors of the
Sh-G model. In a relativistic field theory defined in infinite volume
the elementary Form Factors of a local operator $\mc O$ are the matrix
elements of $\mc O(0,0)$ between the vacuum and a set of $n$-particle
asymptotic states:
\be{FFdef}
F_n^{\mc O}(\th_1,\th_2,\dots,\th_n)=\FF{\mc O(0,0)}_\text{in}\;.
\ee
The knowledge of all Form Factors of an operator is enough for
realizing how it acts on any state of the theory. In fact, a generic
matrix element of the operator can be expressed in terms of its Form
Factors by using the translation operator and the crossing symmetry
which is implemented by an analytic continuation in the rapidity
variables \cite{footnote2}:
\be{cross}
\3pt{\th'_1,\dots,\th'_n}{\mc O(0,0)}{\th_1,\dots,\th_k}=
F_{n+k}^{\mc O}(\th'_1+i\pi,\dots,\th'_n+i\pi,\th_1,\dots,\th_k)\,.
\ee

The Form Factors satisfy a set of functional and recursive equations,
which for integrable models makes it possible to find in many cases
their explicit expressions (for a review, see \cite{smirnov}). For a
scalar operator unitary and crossing symmetry dictate the following
functional equations
\bes
\begin{align}
F_n(\th_1,\dots,\th_k,\th_{k+1},\dots,\th_n)&=S(\th_k-\th_{k+1})\,F_n(\th_1,\dots,\th_{k+1},\th_k,\dots,\th_n)\;, \\
F_n(\th_1+2\pi i,\dots,\th_n)&=F_n(\th_2,\dots,\th_n,\th_1)\;.
\end{align}
\esu
The Form Factors of integrable theories can have two kinds of simple
poles and, except for these singularities, they are analytic in the
strip $0<\mathrm{Im}\,\th_{ij}<2\pi$ (here
$\th_{ij}=\th_i-\th_j$). The first kind of poles corresponds to
kinematical singularities at $\th_{ij}=i\pi$ and their residues give
rise to a set of recursive equations between the $n$-particle and the
$(n+2)$-particle Form Factors
\be{reskin}
F_{n+2}(\th'+i\pi,\th,\th_1,\dots,\th_n)\xrightarrow[\th'\to\th]{} 
\frac{i}{\th'-\th}\left(1-\prod_{j=1}^nS(\th-\th_j)\right)F_n(\th_1,\dots,\th_n)+\dots\;,
\ee
where the dots stand for regular parts.
The second kind of poles is related to the bound states of the theory,
but since there are no bound states in the Sh-G model, there are no
such poles in the Form Factors of this theory and we will not consider
them any further.

Note the striking similarity between the recursive equations
\erf{reskin} and \erf{recurs}: we will show below that, indeed, they
exactly correspond one to the other in the limit $\erf{lim}$. These
recursive equations, together with the requirement of the correct
asymptotic behavior and of the desired analyticity properties, are the
key tools in finding explicit solutions for the Form Factors. In the
Sh-G model a concise expression is provided by the Form Factor of the
exponential operator \cite{mussardo}
\be{koubek}
F_n(k)=\FF{e^{kg\phi}}=
\frac{\sin(k\pi\alpha)}{\sin(\pi\alpha)}\left(\frac{4\sin(\pi\alpha)}{F_\text{min}(i\pi)}\right)^{\frac{n}2}\det
M_n(k)\prod_{j<l}^n \frac{F_\text{min}(\th_j-\th_l)}{e^{\th_j}+e^{\th_l}}\;.
\ee
Here $k$ is an arbitrary real number, $F_{\text{min}}(\th)$ is the
minimal solution of the Form Factor bootstrap equations and $M_n$ is a
$(n-1)\times(n-1)$ matrix
\be{}
\left[M_n(k)\right]_{j,\,l}=\sigma^{(n)}_{2j-l}\,
\frac{\sin\left((j-l+k)\pi\alpha\right)}{\sin(\pi\alpha)}\;, 
\ee 
where $\sigma^{(n)}_j$ are the elementary symmetric polynomials of the
variables $e^{\th_{j}}$:
\be{}
\sigma^{(n)}_j=\sum_{i_1<\dots<i_j}^n e^{\th_{i_1}}\dots e^{\th_{i_n}}\;.
\ee
We will see that the Form Factors of $e^{kg\phi}$ act as generating
functions for the Form Factors of all the powers of the field $\phi$.

\section{QNLS matrix elements from Sh-G Form Factors}
\label{last}
Before going on with the analysis, it is worth emphasizing the strong
similarities of the QNLS and Sh-G models, in particular, the
similarity of the key equations for the Bethe Ansatz matrix elements
and the Form Factors. Both theories are integrable and they contain a
single type of massive particle without additional bound states. The
pseudo-vacuum, on which the BA states are built, coincides with the
Fock-vacuum, i.e. the zero-particle state. The Sh-G Form Factors are
built on the physical vacuum, but in the non-relativistic limit the
zero-point fluctuations disappear and we obtain the Fock-vacuum of the
QNLS model. The two-particle $S$-matrix, which in the relativistic
context governs every aspect of integrable theories, also corresponds
between the two models. From this it is clear that the Bethe equations
\erf{BYlog},\erf{BYrel} are also mapped to each other. The striking
similarity between the recursive equations \erf{recurs} and
\erf{reskin} is very important, because these are the key equations
that allow for the determination of the Form Factors and the matrix
elements. Moreover, the building block of the relativistic Form
Factors, $F_{\text{min}}(\th)$ has the following behavior: in the
limit \erf{lim}
\be{Fminlim}
F_{\text{min}}(\th_{jl})\to\frac{\l_j-\l_l}{\l_j-\l_l+ic}=f(\l_j,\l_l)\;,\qquad
F_{\text{min}}(\th_{jl}+i\pi)\to1 \qquad \text{for }
\th_{jl}\in\mathbb{R}\;, 
\ee 
thus it goes into an important building block of the Bethe Ansatz
quantities. With all these correspondences it should be quite clear
that the Sh-G Form Factors will be mapped to the QNLS matrix
elements. In what follows we give the details of this mapping and
provide explicit examples.

In order to obtain QNLS matrix elements from the Sh-G Form Factors
using the limit \erf{lim}, we have to understand both the relation
between the operators and the states of the two models. We also have
to take into account that the Bethe Ansatz matrix elements are defined
in finite volume whereas the relativistic ones discussed so far are
defined in infinite volume. The latter difference can be cured using
the results of \cite{balazs}.  Even in relativistic quantum field
theories the states in a finite volume $L$ can be described as
multi-particle scattering states, where the rapidities of the
particles are solutions of the Bethe quantization conditions:
\be{BYrel}
Q_j = mc_lL\sinh\th_j+\sum_{k\ne j}^n\frac1i \log S(\th_j-\th_k) =
2\pi I_j\;,\qquad\quad j=1,\dots,n\;.
\ee
However, this procedure is to be understood as an ``asymptotic Bethe
Ansatz" in the sense that it gives the correct results to all orders
in $1/L$ but it misses residual finite size effects which decay
exponentially with the volume. These corrections can be associated to
processes involving virtual particles with some of them ``travelling
around the world". The main idea of \cite{balazs} is that this picture
of ``asymptotic BA" applies to the form factors as well: the matrix
elements in a finite volume are given by the infinite volume Form
Factors at the particular set of rapidities which solve the
corresponding Bethe equations. In addition, one has to introduce
normalization factors given by the corresponding Gaudin determinants
\be{rho}
\rho_n(\th_1,\dots,\th_n)=\det \frac{\p Q_j}{\p\th_k}
\ee
which can be interpreted as the density of states (in rapidity-space)
in the corresponding sector, or alternatively as the norms of the BA
states like in \erf{norm}.  Thus the {\em finite volume} Form Factors
are given by the infinite volume ones taken at the rapidities
satisfying the Bethe equations \erf{BYrel}, divided by the density of
states:
\be{FF_L}
\3pt{\th'_1,\dots,\th'_l}{\mc O(0,0)}{\th_1,\dots,\th_n}_L = 
\frac{F^{\mc O}(\th'_1+i\pi,\dots,\th'_l+i\pi,\th_1,\dots,\th_n)}
{\sqrt{\rho_l(\th'_1,\dots,\th'_l)}\sqrt{\rho_n(\th_1,\dots,\th_n)}}
+\mc O(e^{-\mu L}) \;,
\ee
where $\mc O(e^{-\mu L})$ stands for the above-mentioned exponentially
small corrections in $L$. In the non-relativistic limit both the Bethe
equations and the norms of states go over to the QNLS quantities
\erf{BYlog}, \erf{rhoNR}. In particular, the relation between the norms
is given by
\be{rhos}
\rho_N \longrightarrow (mc_l)^N\tilde\rho_N\;.
\ee
We note that \eqref{FF_L} is valid as long as there are no coinciding
rapidities in the ``bra'' and ``ket'' vectors. In the latter case
disconnected terms arise; a proper treatment of such contributions was
given in \cite{fftcsa2}. In this paper we do not elaborate on
diconnected pieces, i.e. we assume that the two sets of rapidities are
completely distinct.

Let us turn now to the relation between the operators in the two
theories which was given in \cite{KMT1}:
\be{phi-psi}
\phi(x,t)\sim\sqrt{\frac{\hbar^2}{2m}}\left(\psi(x,t)\,
  e^{-i\frac{mc_l^2}\hbar\,t}+\psid(x,t) e^{+i\frac{mc_l^2}\hbar\,t}\right)\,. 
\ee
The exponential terms have to be separated, because the relativistic
Hamiltonian contains also the rest energy which is absent in the
non-relativistic case. The sign $\sim$ means that in any functional
expression of $\phi$ the surviving exponential terms should be
dropped, because in the non-relativistic limit $c_l\to\infty$ they are
rapidly oscillating and give zero when integrated over any small but
finite time interval.

Similarly, we must compensate for the rest energy in the time
evolution of the states, so (without the proper normalizations)
\be{states1}
\vec{\th_1,\dots,\th_n}\;\longleftrightarrow\; 
e^{-in\,mc_l^2 t}\vec{\l_1,\dots,\l_n}\;,
\qquad\quad \cev{\th_1,\dots,\th_n}\;\longleftrightarrow\; 
e^{+in\,mc_l^2 t}\cev{\l_1,\dots,\l_n}\;.
\ee
The relation between the rapidities and the momenta in the
non-relativistic limit is $\l_i=\th_i/mc_l$. Note that the encounter
of the oscillating terms in the states \erf{states} and in the
operator \erf{phi-psi} (as well as in all its powers) will ensure that
in the non-relativistic limit a given operator (e.g. $\psi$) have
non-zero matrix elements only between states in which the number of
particles differ by a fixed amount (e.g. one).

Let us now deal with the question of the normalization of the
states. First of all, one should note that in contrast with the BA
states the relativistic asymptotic states are not symmetric in the
rapidities, they rather obey
\be{}
\vec{\th_1,\dots,\th_k,\th_{k+1},\dots,\th_n}=
S(\th_k-\th_{k+1})\,\vec{\th_1,\dots,\th_{k+1},\th_k,\dots,\th_n}\;. 
\ee
Hence, in order to establish the correspondence with the BA states,
these states should be symmetrized in the rapidities, which can be
done by multiplying with the appropriate phase factors:
\be{phase}
\vec{\th_1,\dots,\th_n}_{\text{symm}}=
\prod_{j>k}\sqrt{S(\th_j-\th_k)}\,\vec{\th_1,\dots,\th_n}\;. 
\ee
The normalization \erf{norm} of the BA states should also be taken
into account. From \erf{rhos} it is clear that for the proper
normalization we have to include a factor of $\sqrt{mc_l}$ for every
particle.

Collecting everything we finally arrive at the relation
\be{}
\prod_{j>k}\sqrt{S(\th_j-\th_k)}\,\vec{\th_1,\dots,\th_n}\;\;\sim\;\; 
\frac{(mc_l)^{n/2}}{c^{n/2}\prod\limits_{j,k=1\atop{j\ne k}}^N
\sqrt{f(\l_j,\l_k)}} \;e^{-in\,mc_l^2 t} \vec{\l_1,\dots,\l_n}\;.
\ee
Taking the S-matrices to the right hand side and using eqn. \erf{SLL}
for the $S$-matrix recovered in the double limit of the
Sh-G model, this can be written as
\be{states}
\vec{\th_1,\dots,\th_n}\;\;\sim\;\; 
\frac{ (mc_l)^{n/2}e^{-in\,mc_l^2 t}}{c^{n/2}\prod\limits_{j<k}^Nf(\l_j,\l_k)}\;\vec{\l_1,\dots,\l_n}\;,
\ee
The relation between the dual vectors is given by the complex
conjugate expression, in particular, the sign of the time-dependent
phase \erf{states1} is opposite.

Using the $S$-matrix \erf{SLL} of the QNLS model and the
correspondence of the states \erf{states} it is easy to prove, as
shown below, that the relativistic recursive equations \erf{reskin}
transform exactly into the recursive equations \erf{recurs}. Note also
that due to \erf{Fminlim} the $f(\l_i,\l_j)$ factors in \erf{states}
will exactly cancel the limiting forms of $F_{\text{min}}(\th_{ij})$.

\bigskip

Now we are in a position to obtain a {\em generic QNLS matrix element} of the form
\be{}
\3pt{\l'_1,\dots,\l'_{N+p-q}}{\psid\,^p\psi^q}{\l_1,\dots,\l_N}
\ee
by performing the following steps:
\begin{enumerate}
\item We determine the infinite volume Sh-G Form Factor 
  \[\3pt{0}{\no{\phi^{p+q}}}{\th'_1+i\pi,\dots,\th'_{N+p-q}+i\pi,\th_1,\dots,\th_N}\]
  by picking the $\mc O(k^{p+q})$ term in the expansion of
  \erf{koubek} and (for $p+q>2$) by taking into account the normal
  ordering issues \cite{footnote3}. The pre-factor in \erf{phi-psi} is
  1 with the choice $m=1/2,\hbar=1$, but simple combinatorial factors
  may arise. As we said, the number of crossed rapidities will {\em
    automatically} select the correct combination $\psid\,^p\psi^q$
  out of $\phi^{p+q}$.
\item We calculate the finite volume Form Factors using
  \erf{FF_L}. However, in the double limit the norms of the
    finite volume states will go to those of the BA states \erf{norm}
    up to a factor of $\sqrt{mc_l}$ for every particle, which needs to
    included.
\item We take the double scaling limit \erf{lim} using \erf{alpha} as
  well as the relation $\th_i\to\l_i/mc_l$ and the limit of the
  minimal Form Factor $F_{\text{min}}$ \erf{Fminlim}.
\item We take into account the different normalizations
  according to \erf{states}.
\item To get a proper matching with the Bethe Ansatz matrix elements
  we include a factor of $-i$ for each $\psi$ and a $+i$ for each $\psid$.
\end{enumerate}
These steps can be summarized in the formula
\begin{multline}
\label{final}
\3pt{\l'_1,\dots,\l'_{N+p-q}}{\psid\,^p\psi^q}{\l_1,\dots,\l_N} \;=\;\\
i^{p-q}\binom{p+q}{p}^{-1} \left(\frac{2m}{\hbar^2}\right)^{\frac{p+q}2}
\;c^{\frac{2N+p-q}2}\prod\limits_{j<k}^Nf(\l_j,\l_k)\prod\limits_{j<k}^{N+p-q}f(\l'_j,\l'_k)\;\times\\
\times\;\widetilde\lim\left\{ (mc_l)^{-\frac{2N+p-q}2}
\3pt{0}{\no{\phi^{p+q}}}{\th'_1+i\pi,\dots,\th'_{N+p-q}+i\pi,\th_1,\dots,\th_N}\right\}\;,
\end{multline}
where $\widetilde\lim$ denotes the double scaling limit
\erf{lim}.

As a first check of this procedure let us explicitly show the
correspondence of the recursive equations \erf{reskin} and
\erf{recurs}. Starting from the relativistic case we have
\begin{multline}
F_{2N-1}(\th'_1+i\pi,\dots,\th'_{N-1}+i\pi,\th_1,\dots,\th_N)=\\
\prod_{j=2}^{N-1}S(\th'_j+i\pi-\th_1)F_{2N-1}(\th'_1+i\pi,\th_1,\th'_2+i\pi,\dots,\th'_{N-1}+i\pi,\th_2,\dots,\th_N) 
\;\;\xrightarrow[\;\;\th'_1\to\th_1\;\;]{}\\
\frac{i}{\th'_1-\th_1}\left\{1-\prod_{j=2}^{N-1}S(\th'_1-\th'_j-i\pi)\prod_{j=2}^{N}S(\th_1-\th_j)\right\}
\prod_{j=2}^{N-1}S(\th'_j+i\pi-\th_1)
F_{2N-3}(\th'_2+i\pi,\dots,\th'_{N-1}+i\pi,\th_2,\dots,\th_N)=\\
\frac{i}{\th'_1-\th_1}\left\{\prod_{j=2}^{N-1}S(\th'_1-\th'_j)-\prod_{j=2}^{N}S(\th_1-\th_j)\right\}
 F_{2N-3}(\th'_2+i\pi,\dots,\th'_{N-1}+i\pi,\th_2,\dots,\th_N)\;.
\end{multline}
In the limit using \erf{final} this relation becomes
\begin{multline}
i\left(\frac{mc_l}{c}\right)^{\frac{2N-1}2}\sqrt{\frac{\hbar^2}{2m}}
\frac1{\prod\limits_{j<k}^Nf(\l_j,\l_k)\prod\limits_{j<k}^{N-1}f(\l'_j,\l'_k)}
 \tilde F_{N}(\l'_1,\dots,\l'_{N-1},\l_1,\dots,\l_N)
\;\;\xrightarrow[\;\;\l'_1\to\l_1\;\;]{}\\
\shoveleft{\frac{i\,mc_l}{\l'_1-\l_1}\left\{\prod_{j=2}^{N-1}
\tilde S_{\text{QNLS}}(\l'_1-\l'_j)-\prod_{j=2}^{N}\tilde S_{\text{QNLS}}(\l_1-\l_j)\right\}\;\times}\\
\times\;i\left(\frac{mc_l}{c}\right)^{\frac{2N-3}2}\sqrt{\frac{\hbar^2}{2m}}
\frac1{\prod\limits_{2<j<k}^Nf(\l_j,\l_k)\prod\limits_{2<j<k}^{N-1}f(\l'_j,\l'_k)}
 \tilde F_{N-1}(\l'_2,\dots,\l'_{N-1},\l_2,\dots,\l_N)\;.
\end{multline}
Simplifying with the common pre-factors and using relation \erf{SLL} between
the $S$-matrix and the function $f(\l,\mu)$ we arrive at equation \erf{recurs}.

\bigskip

Let us take now some explicit examples of the limit for the matrix
elements of the field operator $\psi$ and the current operator
$\psid\psi$.  For the matrix elements of $\psi$ a nice determinant
representation was found in \cite{kojima}. They can be expressed also
as \cite{izkorresh}
\be{}
\tilde F_N^{\psi}(\l'_1,\dots,\l'_{N-1}|\l_1,\dots,\l_N)=
\frac{P_N(\l'_1,\dots,\l'_{N-1}|\l_1,\dots,\l_N)}
{\prod\limits_{k=1}^{N-1}\prod\limits_{j=1}^N(\l_j-\l'_k)}\;,
\ee 
where $P_N$ are polynomials in $\{\l\}$. The first few examples are
given by
\bes
\label{psiFF}
\begin{align}
P_1&=-i\sqrt{c}\;,\\
P_2&=-2i\sqrt{c}\,c^2\;,\\
P_3&=-4i\sqrt{c}\,c^4\left[-c^2+\left((\l_1-\l_1')(\l_2-\l'_2)+
(\l_1-\l_2')(\l_3-\l'_1)+(\l_2-\l_1')(\l_3-\l'_2)\right)\right]\;.
\end{align}
\esu
The form factors for $\psid\psi$ can be extracted from the matrix
elements of the non-local operator
\be{Q1}
Q_1(x)=\int_0^x\ud y\,j(y)
\ee
by differentiating with respect to $x$. Matrix elements of $Q_1(x)$
are listed for example in \cite{izkor} and a determinant formula can
be found in \cite{slavnov}, which yield
\bes
\label{jFF}
\begin{align}
\tilde F_1^{j}(\l'_1|\l_1)&=c\;,\\
\tilde F_2^{j}(\l'_1,\l'_2|\l_1,\l_2)&=-2c^3\frac{(\l_1+\l_2-\l'_1-\l'_2)^2}{\prod\limits_{j,k=1}^2(\l'_j-\l_k)}\;,
\end{align}
\esu
and so on.

On the Sh-G side the Form Factors of the various powers of $\phi$ can
be obtained by series expanding formula \erf{koubek} in the auxiliary
real variable $k$ \cite{footnote3}. For example, the Form Factors of
$\phi$ are given by
\be{FFphi}
\FF{\phi}=
\frac{\pi\alpha}g\left(\frac4{F_{\text{min}}(i\pi)}\right)^{\frac{n}2}(\sin(\pi\alpha))^{\frac{n}2-1}
  \det M_n(0)\prod_{j<l}^n\frac{F_\text{min}(\th_j-\th_l)}{e^{\th_j}+e^{\th_l}}\,,
\ee
or explicitly
\begin{align}
\3pt{0}{\phi}{\th_1} &= \frac2{\sqrt{F_{\text{min}}(i\pi)}}\,
\frac{\pi\alpha}{g\,\sqrt{\sin(\pi\alpha)}}\;,\\
\3pt{0}{\phi}{\th_1,\th_2,\th_3} &=
\frac8{F_{\text{min}}(i\pi)^{\frac32}}\,\frac{\pi\alpha}g\sqrt{\sin(\pi\alpha)}
\,e^{\th_1+\th_2+\th_3}\prod_{j<l}^3\frac{F_{\text{min}}(\th_j-\th_l)}{e^{\th_j}+e^{\th_l}}\;.
\end{align}
Now we can apply the rule \erf{final} with $p=0$, $q=1$ and
$N=1,2,\dots$ It is useful to note that
\be{}
c_l\,\alpha\to\frac{2c}{\pi\hbar}\;,\qquad \frac{\pi\alpha}{g}\to\frac{\sqrt{c}}{2\pi}\;.
\ee
From \erf{Fminlim} we see that $F_{\text{min}}(i\pi)\to1$ and that the
surviving $F_{\text{min}}$ factors exactly cancel the $f$-functions
appearing in \erf{final}. Taking care of the pre-factors the double
limit yields a final result which coincides exactly with the
expressions \erf{psiFF}.

Similarly, the first Form Factors of $\phi^2$ obtained from
\erf{koubek} are 
\bes
\label{hopp}
\begin{align}
\3pt{0}{\phi^2}{\th_1,\th_2}&=\frac8{F_{\text{min}}(i\pi)}\frac{\pi^2\alpha^2}
{g^2\,\sin(\pi\alpha)}\,F_\text{min}(\th_1-\th_2)\;,\\
\3pt{0}{\phi^2}{\th_1,\th_2,\th_3,\th_4}&=
\frac{32}{F_{\text{min}}(i\pi)^2}\frac{\pi^2\alpha^2}{g^2}
\,\prod_{j<l}^4\frac{F_\text{min}(\th_j-\th_l)}{e^{\th_j}+e^{\th_l}}\;\times\\
&\!\!\!\!\!\!\!\!\!\!\!\!\!\!\!\!\!\!\!\!\!\!\!\!\!\!\!\!\times
\left(e^{\th_1+\th_2+\th_3+\th_4}(e^{\th_1}+e^{\th_2}+e^{\th_3}+e^{\th_4})^2-
e^{2(\th_1+\th_2+\th_3+\th_4)}(e^{-\th_1}+e^{-\th_2}+e^{-\th_3}+e^{-\th_4})^2\right)\;.
\end{align}
\esu
Applying the double limit formula \erf{final} with $p=q=1$ to
\erf{hopp} one retrieves the matrix elements \erf{jFF}. We have also
checked the matrix elements with higher number of particles finding a
perfect agreement with the corresponding Bethe Ansatz matrix elements.

\section{Conclusions}

In this paper we have analyzed the close correspondence between matrix
elements of Bethe Ansatz models and Form Factors of relativistic
integrable field theories. If the former models can be regarded as
non-relativistic limits of the latter theories, then the Bethe Ansatz
matrix elements can be efficiently obtained as the non-relativistic
expressions of the corresponding Form Factors, whose explicit
computation is much simpler.

We have discussed in detail this correspondence between the Quantum
Non-Linear Schr\"odinger model and the Sinh--Gordon model, where we
gave a universal method to compute all the matrix elements of every
local operator, but there are strong arguments in favor of its
validity for other pairs of models as well. Indeed, both in Bethe
Ansatz models and quantum field theories the main properties of matrix
elements are dictated by their $S$-matrices: therefore, if there is a
mapping between the Hilbert spaces and operators of the two theories
and the expressions of the two $S$-matrices coincide in the
non-relativistic limit of the quantum field theory, these two facts
induce a mapping between the matrix elements of the two theories. The
use of Form Factors may help in solving most of the technical
obstacles that have prevented so far the computation of matrix
elements in non-relativistic Bethe Ansatz integrable models, thus
opening new perspectives on the computation of their correlation
functions. Along this direction it would be interesting to investigate
multi-component systems, where the BA results are limited due to the
nested nature of Bethe Ansatz.

\vspace{3mm} 
{\bf Acknowledgements} We thank G\'abor Tak\'acs and in particular
Jean-S\'ebastien Caux for useful discussions. M. K. and G. M. were
supported by the grants INSTANS (from ESF) and 2007JHLPEZ (from MIUR).


\begin{thebibliography}{99}
\bibitem{footnote1} 
Matrix elements of the Bethe Ansatz models are also called ``form
factors" in the literature. Although the main conclusion of this paper
is that they are proper limit expressions of the relativistic form
factors, to avoid confusion about which quantities we are referring to
in various parts of the paper, we will call the Bethe Ansatz
quantities ``matrix elements" while we will reserve the term ``Form
Factors" for those relative to quantum field theories.
\bibitem{footnote2} 
If $\th'_i=\th_j$ for some $i$ and $j$, this formula gets modified by
contact terms.
\bibitem{footnote3} 
We note that the Form Factors of normal ordered operators
$\no{\phi^n}$ for $n>2$ are obtained as linear combinations of terms
coming from the expansion of the exponential; we refer the reader to
paper \cite{KMT2} for a detailed discussion of this technical
issue.
\end{thebibliography}
\end{document}